\documentclass[floats,floatfix,showpacs,amssymb,prd,twocolumn,superscriptaddress,nofootinbib,nolongbibliography,reprint]{revtex4-1}

\usepackage{amssymb,amsmath,verbatim,mathtools,needspace,enumitem,etoolbox,graphicx,physics,microtype,afterpage}
\usepackage{gensymb}

\usepackage[dvipsnames, usenames]{xcolor}

\definecolor{linkcolor}{rgb}{0.0,0.3,0.5}
\definecolor{dodgerblue}{HTML}{1E90FF}
\usepackage[unicode, colorlinks=true, linkcolor=linkcolor, citecolor=linkcolor, filecolor=linkcolor,urlcolor=linkcolor, pdfusetitle]{hyperref}
\usepackage[all]{hypcap}
\usepackage[T1]{fontenc}
\usepackage[utf8]{inputenc}
\usepackage{tabularx}
	
\newcommand{\ssim}{\mathchar"5218\relax\,}

\makeatletter
\newcommand*{\balancecolsandclearpage}{\close@column@grid \cleardoublepage \twocolumngrid}
\makeatother

\def\apj{\rm{ApJ}}
\def\apjl{\rm{ApJ}}
\def\apjs{\rm{ApJS}}
\def\aap{\rm{A\&A}}

\def\mnras{\rm{MNRAS}}
\def\prd{\rm{PRD}}

\def\prx{\rm{PRX}}

\def\cqg{\rm{CQG}}
\def\lrr{\rm{LRR}}

\newcommand{\bham}{\affiliation{School of Physics and Astronomy \& Institute for Gravitational Wave Astronomy, \\ University of Birmingham, Birmingham, B15 2TT, United Kingdom}}

\newcommand\orcid[1]{\href{https://orcid.org/#1}{$\!\!$\includegraphics[scale=0.006]{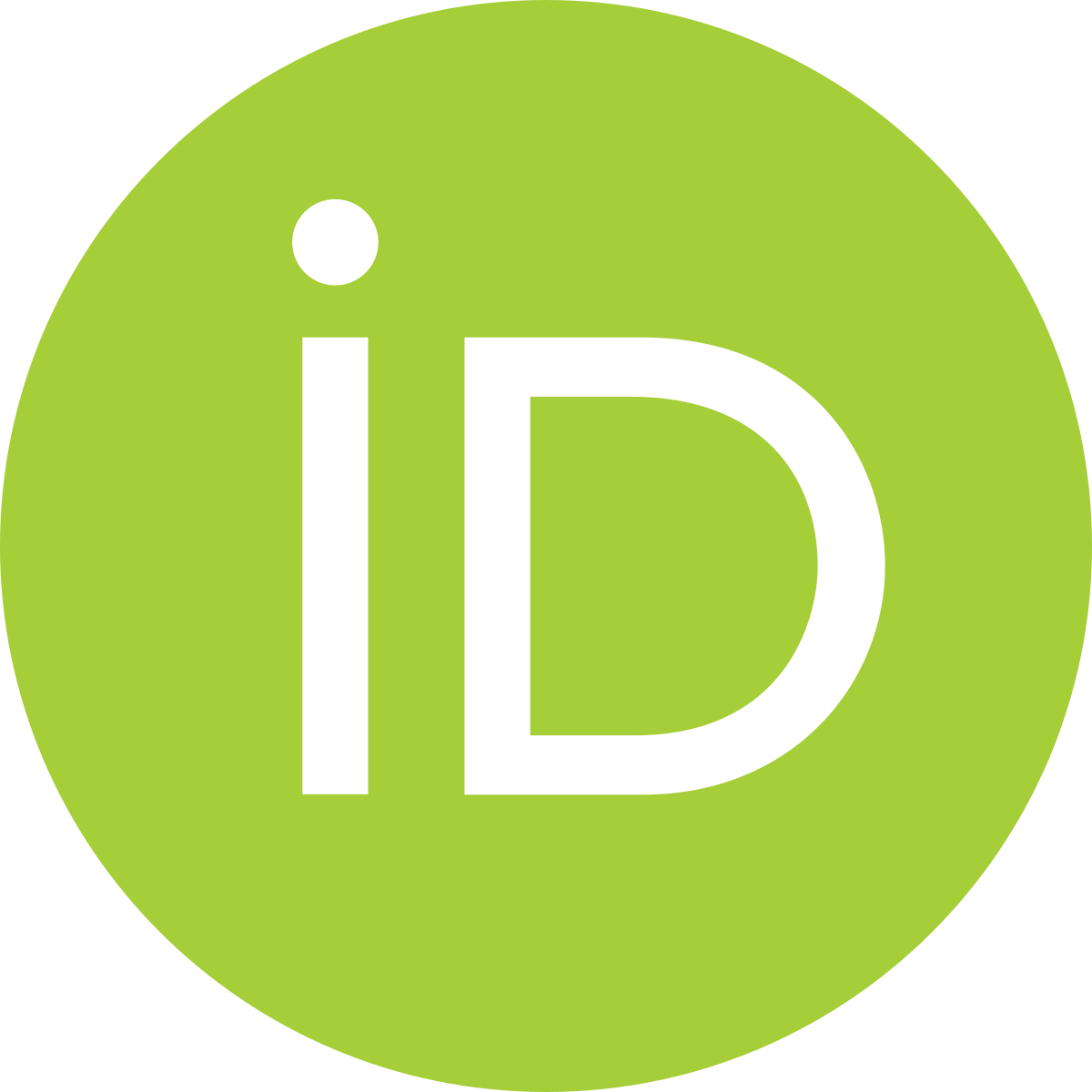} $\!\!$}}

\begin{document}

\title{Gravitational-wave selection effects using neural-network classifiers}

\author{Davide Gerosa \orcid{0000-0002-0933-3579}}
\email{d.gerosa@bham.ac.uk}
 \bham

\author{Geraint Pratten \orcid{0000-0003-4984-0775}}
 \bham

\author{Alberto Vecchio \orcid{0000-0002-6254-1617}}
 \bham

\pacs{}

\date{\today}

\begin{abstract}

We present a novel machine-learning approach to estimate  selection %
effects
in gravitational-wave observations. Using techniques similar to those commonly employed in image classification and pattern recognition, we train a series of neural-network classifiers to predict the LIGO/Virgo detectability of gravitational-wave signals from compact-binary mergers.
We include the effect of spin precession, higher-order modes, and multiple detectors and show that their omission, as it is common in large population studies, tends to overestimate the inferred merger rate in selected regions of the parameter space. Although here we train our classifiers using a simple signal-to-noise ratio threshold, our approach is ready to be used in conjunction with
full pipeline injections, thus paving the way toward including actual distributions of  astrophysical and noise triggers into gravitational-wave population analyses. 

\end{abstract}

\maketitle

\section{Introduction}

Much like any other observatory, gravitational-wave (GW) interferometers suffer from selection biases. Ground-based detectors such as LIGO and Virgo are sensitive only to GW frequencies between $\ssim 10$ and $\ssim 1000$ Hz. While this is the regime of stellar-mass compact objects, the non-flat frequency response of the detectors implies that black holes (BHs) with, say, $50 M_\odot$ are easier to detect than those with, say, $5 M_\odot$. BH binaries with spins that are co- (counter-) aligned with the orbital angular momentum present longer (shorter) signals and are thus easier (harder) to observe. Furthermore, GW detectors are more sensitive to signals coming from specific parts of the sky: a single interferometer is more likely to observe sources if they are located overhead/underneath the detector and presents blind spots in the plane of the arms. This feature is attenuated by a multi-detector network, which in general provides a more uniform sky coverage (although the two LIGO instruments have essentially the same arms' orientation and Virgo is currently still a factor $\ssim2$ less sensitive than LIGO). Finally, because GW emission is beamed along the direction of the binary's orbital angular momentum, observations of face-on sources are far more likely compared to their edge-on counterparts.

As the size of the catalog of observed GW events grows, an accurate characterization of selection effects is  a key ingredient to obtain unbiased estimates of the population properties of merging compact objects. The most accurate way to estimate if a GW signal is observable is to inject a fake copy into the data, which are then processed with the same analysis pipeline employed in the actual search.
Large campaigns with hundreds of thousands of injections are used to characterize the detection efficiency of analysis pipelines, which is a crucial input to infer the astrophysical merger rates of compact objects~\cite{2016PhRvX...6d1015A,2019PhRvX...9c1040A}.
Although accurate, this strategy is computationally very expensive and can be pursued only for a limited number of assumed source populations. However, the population properties and the merger rate depend on each other and must be estimated at the same time. Selection effects thus need to be computed at each evaluation of the population likelihood \cite{2019MNRAS.486.1086M,2019MNRAS.484.4008G,2020arXiv200705579V}, which makes large injection campaigns prohibitive.

In practice, state-of-the-art GW population studies all include selection biases in some approximate form~\cite{2018PhRvD..98h3017T,2019ApJ...882L..24A,2019MNRAS.484.4216R}. The most commonly used approach dates back to Refs.~\cite{1993PhRvD..47.2198F,1996PhRvD..53.2878F} and relies on factoring out the dependencies of the extrinsic angles on the event signal-to-noise ratio (SNR). Events can then be thresholded based on the SNR of an ``optimal'' source with the same intrinsic parameters. %
A recent comparison between outputs of injection campaigns and SNR thresholding has been presented in Ref.~\cite{2016ApJS..227...14A}. Possible paths of improvement include considering additional  calibration factors \cite{2018CQGra..35n5009T,calibratedVT} and time-varying sensitivity noise curves  \cite{2019ApJ...882L..24A}. %

The semi-analytical treatment of Refs.~\cite{1993PhRvD..47.2198F,1996PhRvD..53.2878F} is, strictly speaking, only valid when a single detector ---instead of the full LIGO-Virgo, and soon KAGRA, network--- is considered. Furthermore, the notion of an ``optimal'' source is not straightforward to define for precessing sources (because the orbital plane orientation changes as the binary is observed) and GW emission modes beyond the dominant quadrupole.

In this paper, we present a novel approach to estimate selection effects in GW observations. We tackle this problem with machine learning, borrowing techniques that are commonly applied to image-classification and pattern-recognition problems (see e.g. \cite{GoodBengCour16}). More specifically, we train a series of neural-network classifiers on the parameters of compact binaries, with the goal of predicting if their GW signals are detectable. Can a computer ``learn'' if LIGO and Virgo will observe GWs?

\section{Gravitational-wave detectability}
\label{secgwdet}

A compact binary on a quasi-circular orbit is characterized by source-frame masses $m_1$ and $m_2$ (combined into total mass $M=m_1+m_2$ and mass ratio $q=m_2/m_1\leq 1$), dimensionless spins $\boldsymbol{\chi}_1$ and $\boldsymbol{\chi}_2$, and redshift $z$. Position and orientation with respect to the detectors are defined in terms of right ascension $\alpha$, declination $\delta$, orbital-plane inclination $\iota$, and polarization angle $\psi$.  To simplify the notation, in the following we combine intrinsic and extrinsic parameters using the symbols  $\theta=\{M,q,\boldsymbol{\chi_1},\boldsymbol{\chi_2}\}$ and $\lambda = \{\alpha,\delta,\iota,\psi\}$.

The SNR $\rho$  \cite{1989thyg.book..330T,2009LRR....12....2S} is the most commonly employed metric to estimate the detectability of GW signals. %
For a single interferometer, the SNR $\rho_S$ is defined as the waveform inner product weighted by the detector response and integrated in the frequency domain. The SNR of a network of instruments is given by $\rho_N=\sqrt{\sum_i \rho_{Si}^2}$ where $i$ labels the individual interferometers. In particular, it is common practice to approximate GW selection biases using a threshold in the SNR: an event is deemed as (not) detectable if $\rho>\rho_{\rm thr}$ ($\rho<\rho_{\rm thr}$). Although thresholding the events based on the SNR does not fully take into account the empirical trigger distribution returned by the detection pipelines \cite{2017ApJ...849..118N,2017PhRvD..95d2001M}, it has been shown to faithfully reproduce the performance of current detectors. %
For instance, Refs. \cite{2016ApJS..227...14A, 2014arXiv1409.0522C,2017arXiv170908079C,2018LRR....21....3A} found that the {performance} of the LIGO/Virgo network can be described by either $\rho_S>8$ or $\rho_N>12$. %

Because astrophysical models typically predict masses, spins, and redshifts, one often needs to marginalize the detectability over the extrinsic parameters $\lambda$.  For each value of $\theta$ and $z$, one can define the detection probability as
\begin{equation}
p_{\rm det}(\theta,z)= \int  p(\lambda) \, \Theta[\rho(\theta,z, \lambda) - \rho_{\rm thr}] \, d\lambda\,,
\label{pdetdefine}
\end{equation}
where $p(\lambda)$ is the probability distribution function of $\lambda$ and $\Theta$ indicates the Heaviside step function. This expression is directly related to the so-called effective spacetime volume $VT$ by %
\begin{equation}
VT(\theta)= T_{\rm obs} \int p_{\rm det}(\theta,z) \frac{dV_c}{dz} \frac{1}{1+z}  dz \,,
\end{equation}
where $T_{\rm obs}$ is the time length of the observing run and $V_c$ is the comoving volume.

\begin{figure}
\includegraphics[page=1,width=\columnwidth]{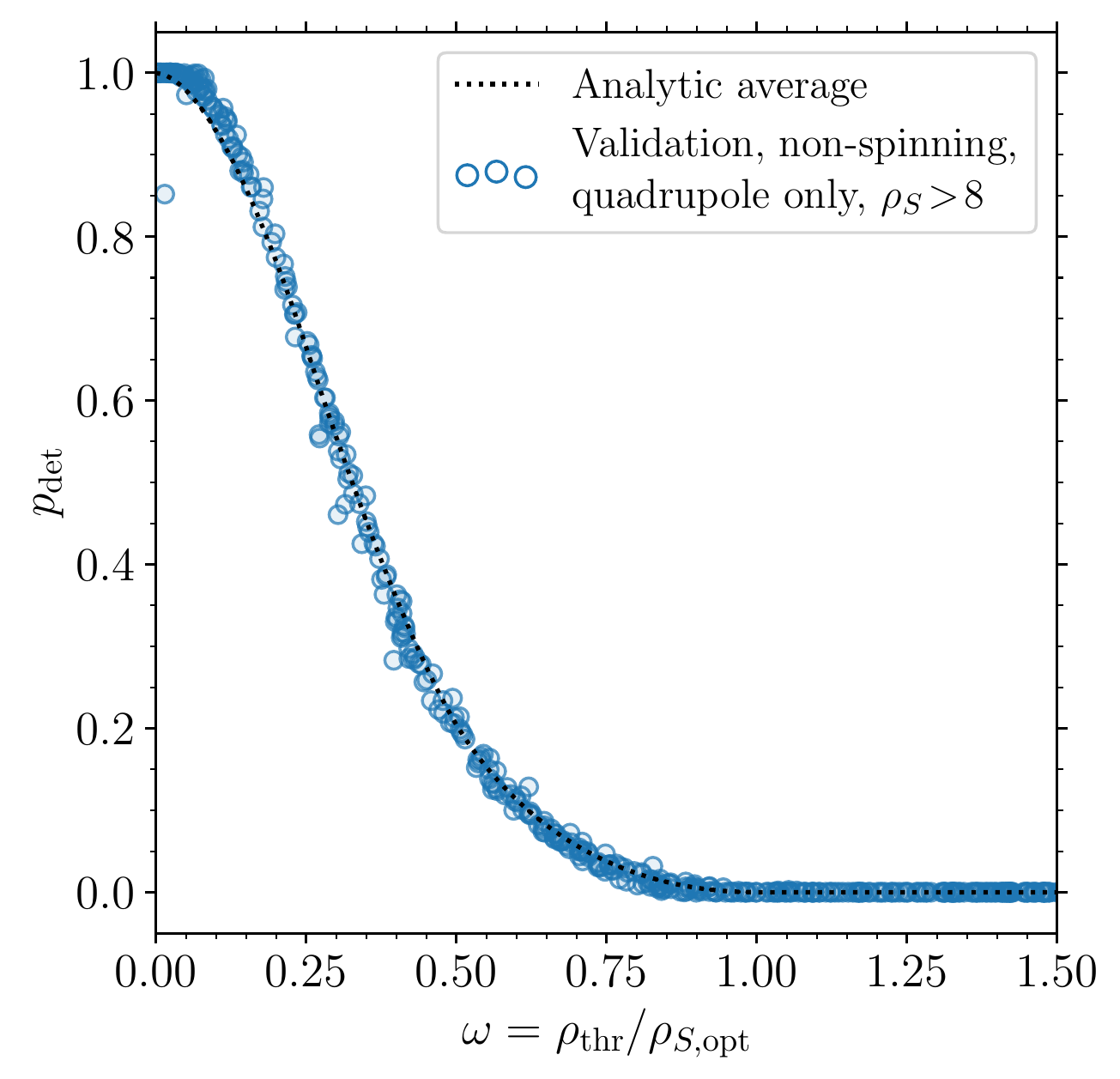}
\label{pdetvsomega}
\caption{Comparison between the semi-analytic and neural-network detectability. The black-dashed line shows the semi-analytic estimate of $p_{\rm det}$ obtained from the cumulative distribution function of the projection parameter $\omega$, evaluated at $\omega = \rho_{\rm thr}/\rho_{S{\rm opt}}$  with $\rho_{\rm thr}=8$. {Trivially, one has $p_{\rm det}=0$ for $\omega>1$.}  The blue circles mark the corresponding estimate obtained by numerically marginalizing over the prediction of our non-spinning neural network evaluated on the validation set. We use $N_{\rm MC}=10^4$ Monte Carlo samples of $p(\lambda)$. A long tail at large values of $\omega$ is omitted for clarity.}
\end{figure}

For the case of a single detector, non-precessing sources, and considering only the dominant quadrupole moment, the  integral in Eq.~(\ref{pdetdefine}) can be computed semi-analytically~\cite{1993PhRvD..47.2198F,1996PhRvD..53.2878F}; the explicit calculation is reported in Appendix~\ref{appd}. In brief, one can factor out  the dependency of $\lambda$ on the SNR to obtain  $\rho_S(\theta,z,\lambda)=\omega(\lambda) \rho_{S,{\rm opt}}(\theta,z)$~\cite{2009LRR....12....2S}, where $\rho_{S,{\rm opt}}$ is the SNR of an ``optimal'' source located  overhead the detector with face-on inclination. Because the projection function $0\leq \omega(\lambda)\leq 1$ is universal,  one can easily convert any given probability distribution function  $p(\lambda)$ into $p(\omega)$. %
Computing the marginalized distribution $p_{\rm det}(\theta,z)$ thus
reduces to evaluating the cumulative distribution function $P(\omega)=\int_\omega^1 p(\omega') d\omega'$ at  $\omega=\rho_{\rm thr}/\rho_{S,{\rm opt}}(\theta,z)$~\cite{2010ApJ...716..615O,2015ApJ...806..263D,2018PhRvD..98h3017T}. %
 The most common application is that of isotropic sources: in this case, $\alpha, \cos\delta, \cos\iota$, and $\psi$ are uniformly distributed  and $P(\omega)$ assumes the familiar shape reported with a dashed black line in Fig.~\ref{pdetvsomega}. In this simplified scenario, estimating GW selection biases requires the evaluation of a single SNR $\rho_{S,{\rm opt}}$.

\section{Machine-learning~implementation}
\label{secmlimpl}

This semi-analytical treatment breaks down in the more general case where a multi-detector network is considered (cf. e.g. Ref.~\cite{2015ApJ...806..263D}). Furthermore, the concept of an optimally oriented source is not directly applicable if one considers precessing binaries and GW harmonics beyond the dominant mode.

We tackle this issue using machine learning. We implement a neural-network classifier using Google's \mbox{TensorFlow} infrastructure \cite{45381}. We use a sequential Keras model with three layers: an input layer with either 7 or 13 neurons (set by the dimensionality of input parameter space, see below), a single hidden layer with 32 neurons, and a final outer layer which classifies the source as either ``detectable'' or ``not detectable.'' The hidden (outer) layer makes use of a hyperbolic tangent (sigmoid) activation function. Inputs are preprocessed with an affine transformation and rescaled within $[-1,1]$ in each dimension.  Neuron weights are initialized using the Glorot algorithm~\cite{glorot2010understanding}.   Optimization is performed using the Adam optimizer \cite{2014arXiv1412.6980K}
with an initial learning rate of $10^{-2}$ and the binary-cross entropy loss function \cite{GoodBengCour16}. The network is exposed to the training data in batches of size 32 
 for up to 150 epochs. 
 After 10 training epochs, the learning rate is decreased exponentially. In general, neural-network performances increase with the size of the training sample, but so does the computational cost of the training. Results presented in this paper are based on neural networks that have been trained and tested on two independent samples of $N=10^7$ sources each.  We systematically explored a wide variety of architectures {varying over number of neurons per layer, number of hidden layers, learning rate, dropout rate, activation function, and batch size. The}  setup we just described has been found to maximize the validation accuracy at  a reasonable computational cost. Training  each of the neural networks  presented in this paper took  $\ssim 30$ hours on a single off-the-shelf CPU. %

We present results obtained with three classifiers. In all cases, training and validation sets are generated distributing detector-frame total masses $M_z=M (1+z)$ uniformly in $[2 M_\odot,1000 M_\odot] $, mass ratios $q$ uniformly in $[0.1,1]$, redshifts uniformly in $[10^{-4}, 4]$, and assuming isotropic orientations $\iota$, sky locations $\alpha, \delta$, and polarization angles $\psi$.
The largest value of $z$ has been chosen to marginally exceed the horizon redshift of all the sources in the sample.
We stress that this distribution does not need to represent a plausible astrophysical scenario but only  allow for accurate training. Once trained, the network can then be evaluated on the chosen population

SNRs are computed with PyCBC  \cite{2016CQGra..33u5004U} assuming the \mbox{IMRPhenomXPHM} \cite{2020arXiv200406503P} %
waveform model and the Planck~15 cosmology \cite{2016A&A...594A..13P}. We consider a three-detector network made of LIGO Hanford, LIGO Livingston, and Virgo at their nominal design sensitivity~\cite{2018LRR....21....3A}. When referring to single-detector SNRs $\rho_S$, we assume a single LIGO instrument. Other  neural networks trained using sensitivity curves of LIGO/Virgo during their observing runs O1, O2, O3, and O4 are provided at Ref.~\cite{pdetclassifier}.

To compare our findings against analytic estimates of $p_{\rm det}$, we first develop a simpler network assuming non-spinning sources ($\boldsymbol{\chi}_1=\boldsymbol{\chi}_2=0$), considering only the dominant $(\ell,|m|) =(2,2)$ mode, and using the single-detector condition $\rho_S>8$. Stepping up in complexity, we then include higher-order modes $(\ell,|m|) =(2,2), (2, 1), (3, 3), (3, 2), (4, 4)$ and precessing sources with spins distributed uniformly in magnitudes in $[0,1]$ and isotropically in directions. In this case, we train networks using both conditions $\rho_S>8$ and $\rho_N>12$.

\begin{figure}
\includegraphics[width=\columnwidth]{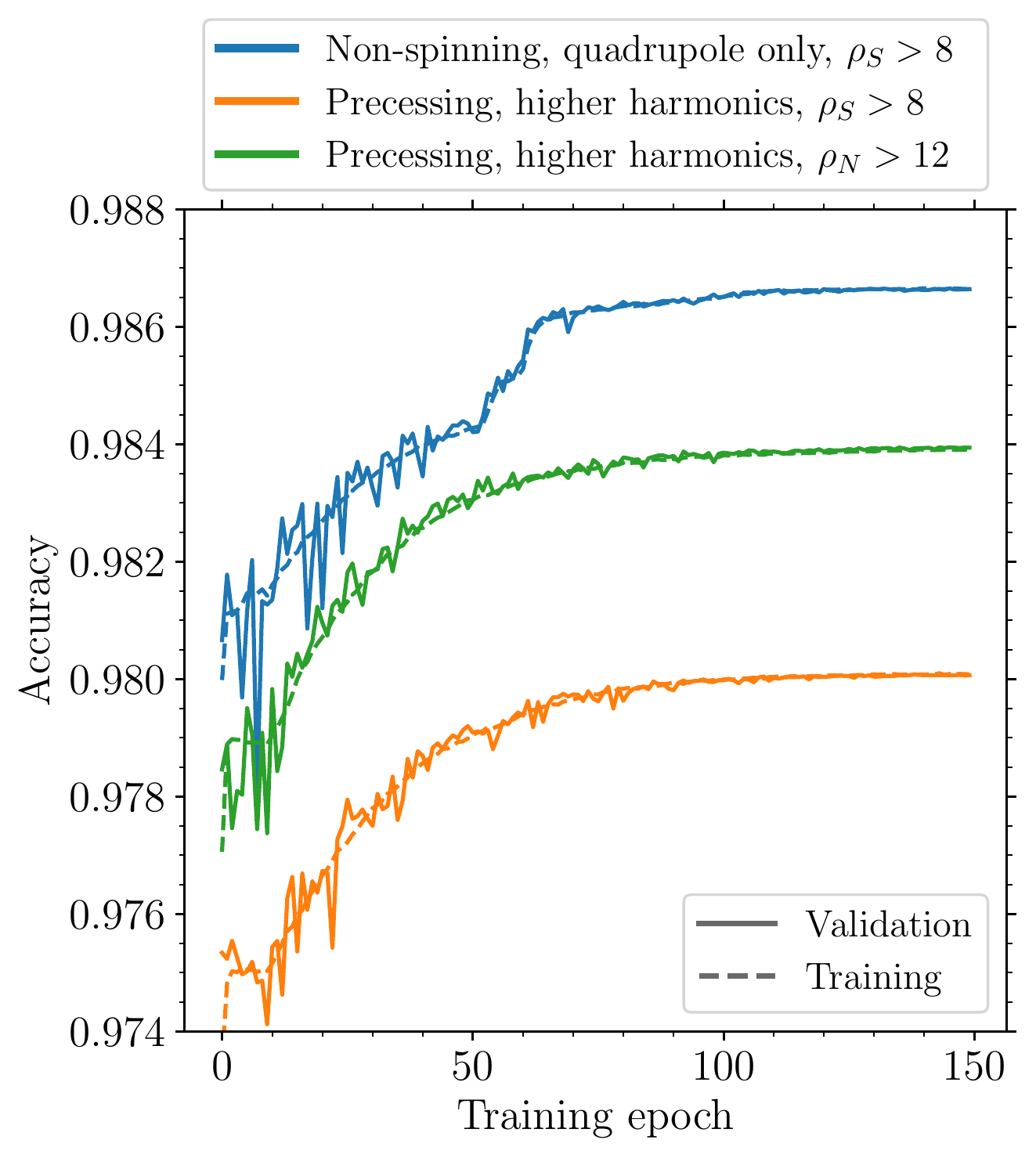}
\caption{Neural-network performances during the training process. Solid (dashed) lines indicate the accuracy evaluated over the validation (training) dataset. Colors indicate three different neural networks trained considering: non-spinning binaries, dominant emission mode, and single detector (blue); precessing binaries, higher-order modes, and single detector (orange); precessing binaries, higher-order modes, and three detectors (green).}
\label{validation}
\end{figure}

\begin{table*}
\setlength{\tabcolsep}{5.5pt}
\begin{tabular}{l|cc|cc|cc|cc}
& \multicolumn{2}{c|}{Training} & \multicolumn{2}{c|}{Validation} & \multicolumn{2}{c|}{``powerlaw''} & 	\multicolumn{2}{c}{``log-uniform''}\\
\hline\hline
& Accuracy & Loss  & Accuracy & Loss &  Accuracy & Loss  & Accuracy & Loss \\
\hline
Non-spinning, quadrupole only, $\rho_S>8$  
& {0.9867} & {0.0320} & {0.9867} & {0.0321}  & $\cdots$ &  $\cdots$ & $\cdots$ & $\cdots$ \\
Precessing, higher harmonics, $\rho_S>8$
&
{0.9801} & {0.0470} & {0.9801} & {0.0471} & {0.9978} & {0.0051} & {0.9921} & {0.0185}
\\
Precessing, higher harmonics, $\rho_N>12$
&
{0.9839} & {0.0383} & {0.9839} & {0.0383} & {0.9981} & {0.0045} & {0.9928} & {0.0172}
\end{tabular}

\caption{Accuracies and losses of the final trained models. We report performances evaluated  on the training and validation sets, which are generated with binaries distributed as specified in Sec.~\ref{secmlimpl}. We also evaluate the classifiers on two astrophysically motivated populations (``powerlaw'' and ``log-uniform'') from Refs.~\cite{2016PhRvX...6d1015A,2019PhRvX...9c1040A}, see Sec.~\ref{secgwpop} for details. All samples in this table have size $N=10^7$. %
}
\label{finaltrained}
\end{table*}

Figure~\ref{validation} shows the evolution of the neural-network accuracies as a function of the training epoch. In this context, accuracy is simply defined as the fraction of the inputs which are correctly identified as either ``detectable'' or ``not detectable''.  Among all iterations, we select the ones that maximize the validation accuracy. The final values of accuracies and losses are reported in Table~\ref{finaltrained}.

As expected, the network trained on non-spinning sources behaves better because it has to interpolate across a smaller 7-dimensional parameter space. When considering the full 13-dimensional parameter space of precessing binaries, we find that the condition $\rho_N>12$ ($\rho_S>8$) is easier (harder) to classify. This is because the beam pattern of a combination of instruments is smoother compared to that of a single interferometer \cite{2011CQGra..28l5023S}. Although more instruments contribute to the SNR, it is worth noting that $\rho_N > 12$ is a potentially stricter criterion than $\rho_S > 8$ with the expected distribution scaling as $\rho^{-4}$~\cite{2011CQGra..28l5023S, 2014arXiv1409.0522C}.  All three neural networks show very similar performances on validation and training sets, indicating that we are not overfitting the data.

Not surprisingly, the events that the network cannot identify correctly all have $\rho \sim \rho_{\rm thr}$. There is roughly an equal number of detectable sources which are classified as ``not detectable'' and vice versa, which suggests that the impact of this mismodeling will be further mitigated when integrating to compute $p_{\rm det}$ and $VT$.

We stress that the loss and accuracy values depend on the population one is trying to predict.
The distributions used in the training/testing process were deliberately chosen to be ``challenging'' to classify: we overpopulate regions of the parameter space at high SNR (low $z$ and high $M$) to encourage the network to better learn the various correlations between the input parameters {for sources with $\rho \sim \rho_{\rm thr}$.}

\section{Gravitational-wave populations}
\label{secgwpop}

\begin{figure}
\includegraphics[page=2,width=\columnwidth]{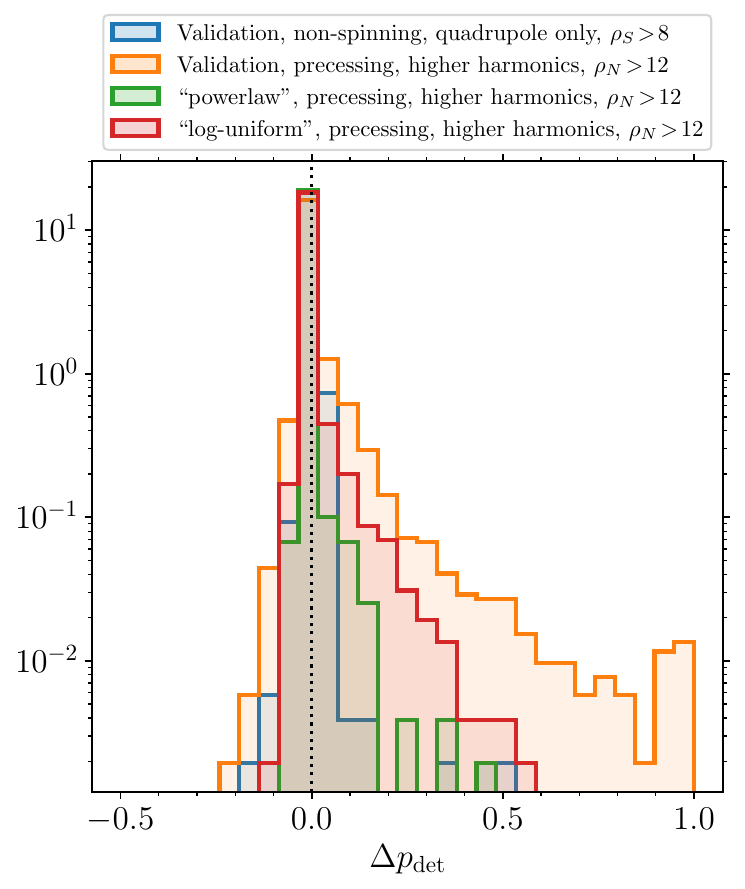}
\caption{Probability distribution of the difference between the detectability $p_{\rm det}$ estimated using our neural networks and the semi-analytic calculation. The latter is only justified in the case of non-precessing sources and a single detector (blue histogram). In the other three cases, the assumptions underlying the analytic calculation are not valid.  Here ``Validation'' (blue and orange histograms) refers to the parameter distributions used in the nominal validation process (cf. Fig.~\ref{validation}) while ``powerlaw'' and ``log-uniform'' (green and red histograms) refer to two test populations similar to those used in Refs.~\cite{2016PhRvX...6d1015A,2019PhRvX...9c1040A} to estimate BH merger rates.}
\label{figpopulation}
\end{figure}

We now compare our results against the analytic estimate of $p_{\rm det}$ described in Sec.~\ref{secgwdet} and Appendix~\ref{appd}.
Once a classifier has been trained, the detectability $p_{\rm det}(\theta,z)$ can be estimated by repeatedly sampling $p(\lambda)$ and counting the number of draws which are identified as ``detectable''.  Figure \ref{pdetvsomega} shows such comparison for the non-spinning, $(2,2)$-mode neural network trained on the condition $\rho_S>8$ and applied to (a fraction of) the test sample. In this case, all the assumptions made in Sec.~\ref{secgwdet} to analytically average $p_{\rm det}$ hold and the two approaches should give the same result.  The blue histogram in Fig.~\ref{figpopulation} shows the deviation $\Delta p_{\rm net}$ between numerical and semi-analytical evaluations. Our neural network reproduces the LIGO detectability both qualitatively and quantitatively. We report minor systematic deviations at large SNRs (small value of $\omega$), where the neural network tends to slightly overestimate the effective spacetime volume (top-left region in Fig. \ref{pdetvsomega}). More suitable network infrastructures and/or training strategies might suppress this spurious feature.

Now that we have established that our approach %
is accurate, we explore whether determining $p_{\rm det}$ using a single-detector SNR threshold for non-precessing binaries and including only the leading order $(2,2)$ mode ---strategy adopted in the overwhelming majority of population studies--- describes  the detection probability correctly across the source parameter space in observations performed with multiple instruments. We therefore 
apply our model to cases where the assumptions behind the semi-analytical estimate of $p_{\rm det}$ are not valid. We evaluate the difference  $\Delta p_{\rm det}$ between the network prediction of $p_{\rm det}$ and the corresponding estimate if one were to naively apply the semi-analytic approach. More precisely, we compute $\omega=8/\rho_{S,{\rm opt}}$ assuming overhead sources with the orbital angular momentum aligned with the line of sight at the GW emission frequency $f_{\rm ref}=20$~Hz. The left (right) region of Fig.~\ref{figpopulation} with $\Delta p_{\rm det}<0$  ($\Delta p_{\rm det}>0$) corresponds to cases where the simplistic approach overestimates (underestimates) the detector performance compared to the prediction of our neural networks.  

\begin{figure}
\includegraphics[width=\columnwidth]{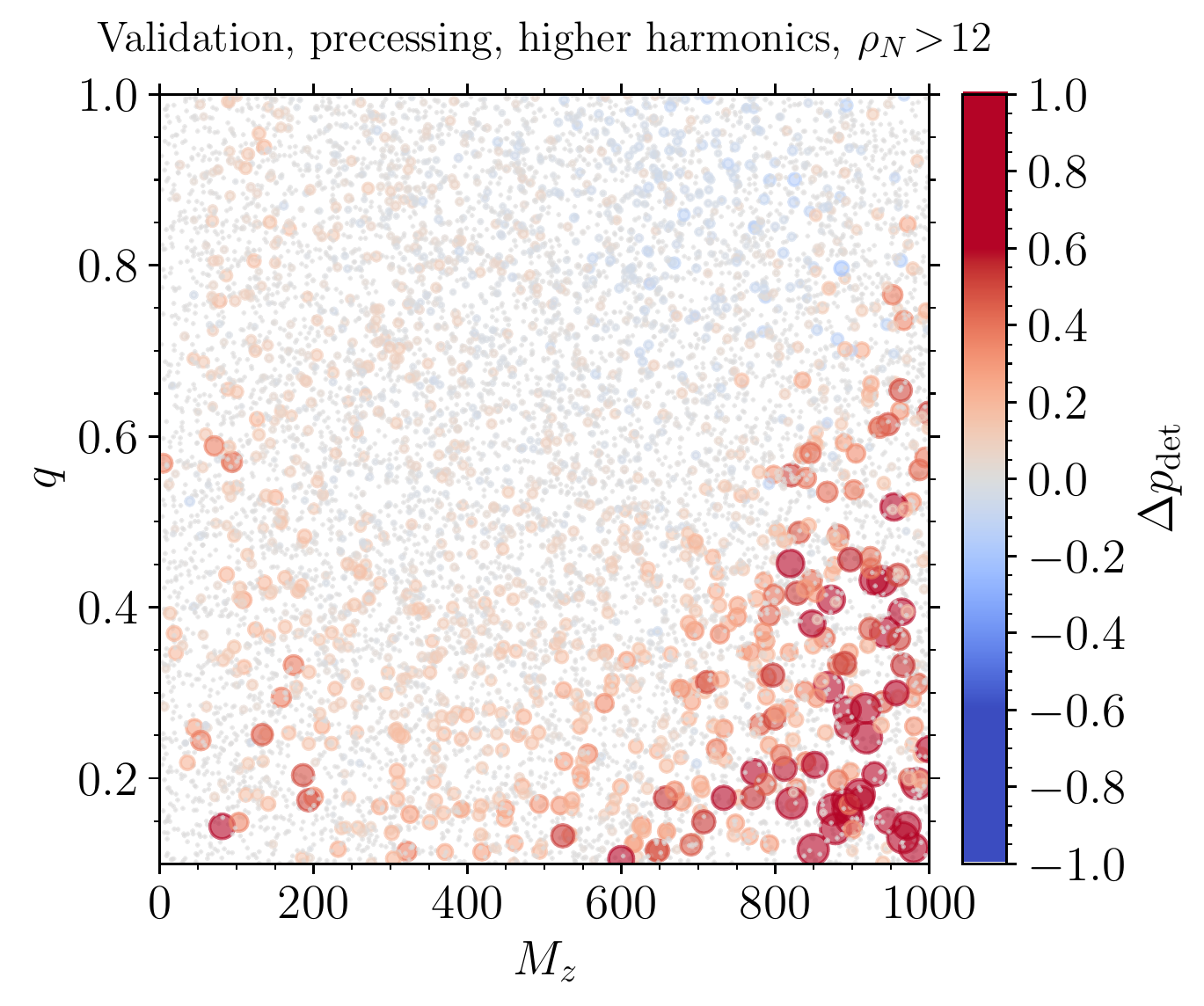}
\caption{Deviation between the numerical and semi-analytical estimates of $p_{\rm det}$ as a function of redshifted mass $M_z$ and mass ratio $q$.  This figure is produced using the validation sample of our model trained on precessing systems, higher harmonics, and the full LIGO/Virgo network. For large values of $M_z$ and low values of $q$, we predict, {on average}, larger value of $p_{\rm det}$ compared to the semi-analytic estimate.}
\label{whichregion}
\end{figure}

The orange histogram in Fig.~\ref{figpopulation} is computed assuming the same distribution used in the  training and validation process, but considering precessing sources, additional harmonics beyond the dominant emission mode, and the three-detector LIGO/Virgo network. Although the vast majority of the sources are correctly identified ($\Delta p_{\rm det}\simeq 0$), the shoulder extending toward positive values of $\Delta p_{\rm det}$ indicates that the analytical single-detector approximation tends to, on average, underestimates the LIGO/Virgo response (see also Refs.~\cite{2015ApJ...806..263D,2018PhRvD..98h3007N,2018PhRvD..98h4036G} for GW selection biases in spinning BH binaries). As shown in Fig.~\ref{whichregion}, the ``culprits'' of these deviations are binaries with large redshifted mass $M_z$ and small mass ratio $q$. This is the region of the parameter space where GW harmonics beyond the dominant  $(\ell,|m|) =(2,2)$ mode provide a significant contribution to the SNR. %

Finally, we estimate the performance of our tool on two plausible astrophysical distributions which are different from those used in the training process. Our populations closely mimic those used in Refs.~\cite{2016PhRvX...6d1015A,2019PhRvX...9c1040A} to estimate merger rates using full software injections. 
In the first scenario (``powerlaw''), we distribute $m_1$ according to $p(m_1)\propto m_1^{-2.3}$ in $[5 M_\odot,100 M_\odot]$ and $m_2$ uniformly in $[5 M_\odot,m_1]$. For the second population (``log-uniform''), we assume that $m_1$ and $m_2$ are distributed uniformly in log: $p(m_1,m_2)\propto 1/m_1 m_2$ with $m_2\leq m_1$. In both cases, we distribute spins uniformly in magnitudes $[0,1]$ and isotropic in directions, and redshifts uniformly in comoving volume and source-frame time, i.e. $p(z)\propto (dV_c/dz) /(1+z)$, in $[10^{-4},4]$. The extrinsic angles $\lambda$ are  isotropically distributed.

Accuracies on these two populations are reported in Table~\ref{finaltrained}. Our neural network can correctly classify more than 99\% of the sources. If one considers only the subsample of sources with $z<1$, the accuracy drops by $\ssim 1\%$ ($\ssim 3\%$) for the ``powerlaw'' (``log-uniform'') case. This is because such a cut has the effect of preferentially selecting sources with $\rho\sim \rho_{\rm thr}$ which are harder to classify.   The green and red histograms in Fig.~\ref{figpopulation} show the corresponding values of $\Delta p_{\rm det}$. The region with $\Delta p_{\rm det}>0$  (to the right of the dashed line in Fig.~\ref{figpopulation}) is more populated and corresponds to cases where our neural networks predicts a larger detection probability compared to the analytic estimate. Our results suggest that neglecting spins, higher-order modes, and considering a single detector (like is done in current population studies) have the net effect of underestimating the effective spacetime volume $VT$, which in turn results in an overestimate of the astrophysical merger rate.

\section{Conclusions}

We presented a new machine-learning approach to estimate selection biases in GW detectors. Borrowing techniques from the field of image recognition, we showed that artificial neural networks can  efficiently be trained to predict the detectability of GW signals emitted by compact binary coalescences. Our predictions appear to be solid both qualitatively and quantitatively. %
 We showed that the inclusion of spins, higher-order modes, and multi-detector SNR calculations results in higher detection probability, suggesting that current detection-rate estimates that do not rely on actual systematic injection studies might be slightly overestimated. Such mismodeling appears to be more relevant in the corner of the parameter space characterized by high and unequal masses. 

Compared to raw evaluations of the SNR, our approach allows for a computational speed-up of more than a factor 1000 (evaluated on a standard off-the-shelf laptop). If a similar number of Monte Carlo samples is used to marginalize over $\lambda$, this implies that our tool (in its present form) might allow the full inclusions of spins, higher-order modes, and network SNRs in $VT$ estimates at the same computational cost of the currently employed semi-analytic average.

We hope our work might be useful to both GW data analysts  and astrophysicists. Researchers developing codes of binary-star and cluster evolution to predict GW events will be able to use our neural network  to efficiently filter their synthetic catalogs based on the LIGO/Virgo detectability. 
Our models are publicly available at \href{https://github.com/dgerosa/pdetclassifier}{github.com/dgerosa/pdetclassifier} \cite{pdetclassifier}, where we provide trained networks and training/validation samples for the three cases described in this paper, as well as additional outputs calibrated on the LIGO/Virgo performances during their observing runs O1, O2, O3, and  forecasted O4. We hope this will facilitate the immediate adoption of our approach.

Our neural-network classifiers are trained to learn a single yes/no variable (``is this input  binary detectable?''). An alternative strategy to estimate GW selection biases using machine learning would be to implement a regression neural network to interpolate the SNR \cite{2020arXiv200710350W}. In this case, the detectability condition $\rho>\rho_{\rm thr}$ can be evaluated at runtime using the predicted values of $\rho$. A classifier like ours has the obvious drawback that a new model needs to be trained for each choice of the SNR threshold $\rho_{\rm thr}$ (but note that the vast majority of the computational cost lies in building the training sample and does not need to be repeated). The advantage is that our  infrastructure can be trained on detectability conditions that are more accurate than a simple SNR threshold. 

Our approach will show its full potential when used in conjunction with large software injections campaigns, like those presented in Refs.~\cite{2016PhRvX...6d1015A,2019PhRvX...9c1040A} to estimate detection rates. Although undoubtedly more accurate at modeling selection biases, such estimates are not currently used in most
GW population studies because of their high computational cost. %
We believe a machine-learning treatment like the one explored in this paper is a promising way forward. 
We recommend that analyses reporting GW %
events should make public %
the estimate of the detection probability across the parameter space covered by the search, to be used down-stream in population studies through, e.g., an approach like ours.

\acknowledgements
We thank Riccardo Buscicchio, Kaze Wong, Patricia Schmidt, Emanuele Berti, Carl-Johan Haster and Eve Chase for discussions.
D.G. is supported by European Union's H2020 ERC Starting Grant No. 945155--GWmining, Leverhulme Trust Grant No. RPG-2019-350, and Royal Society Grant No. RGS-R2-202004. A.V. acknowledges the support of the Royal Society and Wolfson Foundation, and the UK Science and Technology Facilities Council through Grant No. ST/N021702/1. Computational work was performed on the University of Birmingham BlueBEAR cluster, the Athena cluster at HPC Midlands+ funded by EPSRC Grant No. EP/P020232/1 and the Maryland Advanced Research Computing Center (MARCC).

\appendix

\section{Semi-analytic estimate of $p_{\rm det}$}
\label{appd}

The integral in Eq.~(\ref{pdetdefine}) can be carried out semi-analytically for the case a single detector, non-precessing sources, and considering only the dominant quadrupole emission mode. 

In this simpler case, let us define the sky location of the source with a polar angle $\vartheta$ and an azimuthal angle $\phi$. This is equivalent to $\delta=\pi/2 - \vartheta$ and $\phi =\alpha$ for a detector located at the north pole with one arm directed toward the vernal equinox. Given an incoming GW with polarizations $h_+$ and $h_\times$, an interferometer with  $90^\circ$-arms is sensitive to the combination 
 \begin{equation}
h(t) = F_+ h_+(t) + F_\times h_\times(t), 
\label{hoft}
\end{equation}
where the beam patterns are given by \cite{2009LRR....12....2S}
\begin{align}
F_+&= \frac{1}{2}\left (1+\cos^2\vartheta \right )\cos 2\phi \cos 2\psi - 
\cos\vartheta\sin 2\phi \sin 2\psi\,, \label{Fplus}
\\
F_\times &=\frac{1}{2}\left (1+\cos^2\vartheta \right )\cos 2\phi \sin 2\psi + 
\cos\vartheta\sin 2\phi \cos 2\psi\,.
\label{Fcross}
\end{align}
If spin precession and higher harmonics are neglected, the GW emission of  binary coalescence is given by
\begin{align}
h_+(t) &= A(t) \frac{1+\cos^2\iota}{2} \cos\Phi(t), 
\label{hoftp}
\\
h_\times(t) &= A(t)  \cos\iota \sin\Phi(t),
\label{hoftc}
\end{align}
where amplitude $A(t)$ and phase $\Phi(t)$ are predicted by the chosen waveform approximant.
One can rewrite Eq.~(\ref{hoft}) as   
\begin{equation}
h(t) = \omega \, A(t) \, \cos[\Phi(t)- \Phi_0]\,.
\label{factorization}
\end{equation}
where \cite{1993PhRvD..47.2198F,1996PhRvD..53.2878F}
\begin{align}
\omega &= \sqrt{ \left(F_+  \frac{1+\cos^2\iota}{2} \right)^2 + \left( F_\times \cos\iota \right)^2}\,,
\\
\tan \Phi_0 &= \frac{2  F_\times \cos\iota }{ F_+(1+\cos^2\iota)}\,.
\end{align}
It is straightforward to show that  $\max_{\iota,\vartheta,\phi,\psi} \omega =1$.  This maximum is obtained when $\iota=0$ (i.e. the source is face-on) and $\vartheta=0$ (i.e. the source is overhead the detector). Let us call such a source ``optimal''.  The factorization of Eq.~(\ref{factorization}) implies that the SNR of generic binary $\rho$ can be written as $\rho = w \rho_{\rm opt}$. One has $p(\lambda) d\lambda=p(\omega)d\omega$ and thus, from  Eq.~(\ref{pdetdefine}),
\begin{align}
p_{\rm det}&= \int   p(\omega) \, \Theta[\omega\rho_{\rm opt} - \rho_{\rm thr}]  \,d\omega \notag\\
&= \int_{\omega\rho_{\rm opt} > \rho_{\rm thr}} 
\!\!\!\!\!\!\!\!\!\!\!\!\!\!\!\!\!\!\!
p(\omega)\,d\omega  
=
\int_{\rho_{\rm thr}/\rho_{\rm opt}}^1  
\!\!\!\!\!\!\!\!\!\!\!\!\!\!
p(\omega) \,d\omega  \,.
\label{finalintegral}
\end{align}
{The above expression is formally valid for $\rho_{\rm opt}\geq \rho_{\rm thr}$. For $\rho_{\rm opt}< \rho_{\rm thr}$ one has, trivially, $p_{\rm det}=0$.}

Let us note that Refs.~\cite{1993PhRvD..47.2198F,1996PhRvD..53.2878F} make use of the equivalent notation $\Theta\equiv4 \omega$.  For the case of isotropic sources, an analytic fit to the integral in Eq.~(\ref{finalintegral}) is provided in Ref.~\cite{2015ApJ...806..263D}; a public Monte Carlo implementation is available at Ref.~\cite{gwdet}.

\bibliography{detectabilitynnets}

\end{document}